\documentclass[superscriptaddress,runinaddress,showpacs,showkeys,aps,prd,floatfix,onecolumn]{revtex4}
\usepackage{amssymb,amsmath,epsfig}
\usepackage{amsmath,graphicx}
\usepackage{palatino}
\usepackage{changes}
\begin{document}

\title{Weak deflection angle by Casimir wormhole using Gauss-bonnet theorem and its shadow}

\author{Wajiha Javed}
\email{wajiha.javed@ue.edu.pk; wajihajaved84@yahoo.com} 
\affiliation{Division of Science and Technology, University of Education, Township-Lahore, Pakistan}

\author{Ali Hamza}
\email{alihamza.ahg@gmail.com} 
\affiliation{Division of Science and Technology, University of Education, Township-Lahore, Pakistan}

\author{Ali {\"O}vg{\"u}n}
\email{ali.ovgun@emu.edu.tr}
\homepage[]{https://aovgun.weebly.com/}

\affiliation{Physics Department, Eastern Mediterranean
University, Famagusta, North Cyprus via Mersin 10, Turkey.}

\date{\today}

\begin{abstract}

In this paper, we calculate the weak deflection angle by Casimir wormhole and its shadow. To do so, we derive the Gaussian optical curvature and use the Gauss-Bonnet theorem. Then we find the deflection angle by Casimir wormhole in weak field limits. Moreover, we obtain the weak deflection angle in the presence of plasma medium and see the effect of the plasma medium on the weak deflection angle. Moreover, we study a shadow of Casimir wormhole and we plot and discuss them. We show the shadow of Casimir wormhole's behaviour when changing the value of $a$.
\end{abstract}

  

\maketitle
  \section{Introduction}

Idea of wormhole was first suggested by Einstein and Rosen in 1935. This is also known as Einstein-Rosen bridges \cite{A1 1}. A wormhole is a shortcut of the spacetime which makes alternate routes between
two particular spacetimes. The idea of "wormhole" was developed firstly in 1957 inside the seminal papers of
Misner and Wheeler \cite{4 1} and Wheeler \cite{4 2}. Afterward, Wheeler demonstrated that wormholes would be unsteady and non-traversable for even a photon \cite{A1 2}. The metric of Ellis wormholes was first discussed in \cite{MPLA1} and then investigation of Ellis wormholes in General Relativity (GR) went back to the basic works of Morris and Thorne in 1988 \cite{4 3} where they presented a traversable wormhole geometry. Furthermore,
some scientists briefly have examined the gravitational lensing of Ellis wormhole in (\cite{MPLA2}-\cite{MPLA9}). Then the weak and strong gravitational lensing of compact objects are studied in (\cite{MPLA10}-\cite{MPLA26}) and the properties of celestial bodies during gravitation lensing and also test theory of gravity are studied in (\cite{MPLA27}-\cite{MPLA36}).
Later on motivated by the Morris-Thorne paper, numerous scientists study wormhole spacetimes (\cite{A2 4}-\cite{RH4}). Then, Matt Visser proposed a thinshell wormhole in a 1989 \cite{A1 4}. Wormholes are also using to explain quantum entanglement \cite{A1 5}. Exotic matter causes trouble for creating stable wormholes. The Casimir effect which is found by Casimir shows that there is a finite energy between the plates and also calculate the finite force between the plates in his famous paper \cite{1}. Casimir energy is indicated as a potential source and it is also used to prove the existence of negative energy which can be built in the laboratory \cite{CW}. Casimir wormhole is a type of traversable wormhole that can be generate by  using Casimir energy.
\\
  In this paper, we calculate the deflection angle of light (DAL) from Casimir wormhole in weak field limits (WFL). In 2008, Gibbons and Werner (GW) suggested a technique to calculate the DAL by different black holes in the WFL \cite{R 35}. For calculating deflection angle, first  we calculate the Gaussian optical metric using optical geometry of the Casimir wormhole and then we apply the Gauss-Bonnet theorem (GBT) \cite{R 35}:
\begin{equation}
    \alpha=-\int \int_{D_{\infty}} \mathcal{K}dS.
\end{equation}

 Different authors use this technique to calculate DAL by various black holes and wormholes spacetime (\cite{R 36}-\cite{R 45}).  Crisnejo
and Gallo analyzed the DAL in the presence
of plasma medium \cite{M1 19}. Some other works can be found in Refs \cite{Ono:2017pie,Jusufi:2017hed,Ishihara:2016sfv,Arakida:2017hrm,Ovgun:2018tua,Ovgun:2018fte,Ovgun:2018prw,Jusufi:2018jof,Ovgun:2018ran,Jusufi:2017uhh,Ovgun:2018fnk,Ono:2018ybw,Ono:2018jrv,Crisnejo:2018ppm,aoli,Li:2019mqw,Li:2019qyb,Li:2019vhp,Ovgun:2018oxk,Javed:2019qyg,19,Javed:2019kon,Ovgun:2019wej,Javed:2019rrg,Jusufi:2017vew,Crisnejo:2019xtp,Javed:2019ynm,deLeon:2019qnp,Kumaran:2019qqp,km2,Ovgun:2019qzc,km3,Ahmedov:2019dja,Turimov:2018ttf,Abdujabbarov:2017pfw,Schee:2017hof}.\\

It is generally accepted that majority of the galaxies have super-massive black holes at their centers \cite{S1 43,S1 44}, for example Milky Way and Messier 87 having super-massive black hole named as Sgr A and M87. The Event Horizon Telescope imaged the shadow of M87* with measured diameter of 42 microarcsecond  \cite{izr4}.  Recently, attempts have been made to directly visualize the shadow of Sagittarius A* \cite{izr4}, the supermassive black hole in the Galactic Center, and likely observational results will soon be revealed. Shadow of Schwarzschild black hole is first studied by Synge \cite{S1 45} and it is additionally investigated by Luminet \cite{S1 46}. Bardeen \cite{S1 47,S1 48} analyzed shadow cast of Kerr black hole and more work was made by Falcke \cite{S1 49}. Some other works can be found in Refs \cite{Kumar:2020hgm,Konoplya:2019xmn,Kumar:2019pjp,Konoplya:2019goy,Allahyari:2019jqz,Amir:2018pcu,Kho1,Vag1,Vag2,Konoplya:2019sns,Younsi:2016azx,Cunha:2019hzj,Ovgun:2020gjz}. \\
 Main purpose of the paper is to calculate the deflection angle of Casimir wormhole in non-plasma and plasma medium and find shadow cast by Casimir wormhole. The paper is organized as follows: In Sect. II, we calculate the Gaussian optical metric of Casimir wormhole and use the GBT method to calculate its deflection angle in Sect. III. In Sect. IV, we observe the graphical behavior of deflection angle in non-plasma. In Sect. V, we calculate the DAL by Casimir wormhole in plasma medium. In Sect. VI, we analyzed the plots of Casimir wormhole in plasma medium, next we find the null geodesic in Sect. VII and in Sect. VIII we find the shadow of Casimir wormhole and finally in Sect. IX, we conclude our result.

\section{Optical metric of casimir wormhole}
The metric of a Casimir traversable wormhole in a spherically symmetric spacetime is define as \cite{CW};
\begin{equation}
    ds^2=-A(r) dt^2+\frac{dr^2}{B(r)}+r^2 d\Omega^2.\label{AH0}
\end{equation}
Where $A(r)$ and $B(r)$ are defined as
\begin{equation}
   A(r)=\left({\frac{3r}{3r+a}}\right)^2 ~,~ B(r)={1-\frac{2a}{3r}-\frac{{a}^2}{3{r^2}}}~and~d\Omega^{2}=
   d\theta^2+\sin^2\theta d\phi^2.\\
\end{equation}
Here $a$ is a constant and $r$ is a radial coordinate such that $r \in[a,\infty]$. After expanding $A(r)$ as a series and take only second order of $a$, we can write A(r) as :
\begin{equation}
    A(r)=1-\frac{2a}{3r}+\frac{{a}^2}{3{r^2}}+\mathcal{O} (\frac{a^3}{r^3}).\\
\end{equation}
We can assume $(\theta=\frac{\pi}{2})$, and find the optical metric by $ds^{2}=0$,
\begin{equation}
    dt^2=\frac{dr^2}{A(r)B(r)}+\frac{r^2d\phi^2}{A(r)} .\label{H1}
\end{equation}
Now by using Eq. (\ref{H1}), the non-zero christofell symbols are defined as
\begin{equation}
    \Gamma^{0}_{00}=-\frac{B'(r)}{2B(r)}-\frac{A'(r)}{2A(r)}~~~~~~~~,~~~~~~~
    \Gamma^{1}_{01}=\frac{1}{r}-\frac{A'(r)}{2A(r)}
\end{equation}
and
\begin{equation}
 \Gamma^{0}_{11}=-rB(r)+\frac{r^{2}A'(r)B(r)}{2A(r)},
\end{equation}
and Ricci scalar related to the optical metric is calculated as:
\begin{eqnarray}
   \mathcal{R}&=&-\frac{A(r)B'(r)}{r}+\frac{A'(r)B'(r)}{2}+
   \frac{A'(r)B(r)}{r}+{A''(r)}{B(r)}\nonumber\\
   &-&\frac{(A'(r))^{2}B(r)}{2A(r)}.
\end{eqnarray}
The Gaussian optical curvature that is calculated as follows:
\begin{equation}
     \mathcal{K}=\frac{RicciScalar}{2}.
\end{equation}
After simplifying, Gaussian optical curvature is given as:
\begin{equation}
    \mathcal{K}\approx -{\frac {2a}{{3r}^{3}}}+\,{\frac {2{a}^{2}}{3{r}^{4}}}+\mathcal{O}  (\frac{a^3}{r^3}).\label{AH1}
\end{equation}
\section{Deflection angle of Casimir wormhole}
Now by using the GBT we can obtain the DAL of Casimir wormhole.
We use the GBT to the area of $\mathcal{E}_{S}$, given as \cite{R 35}
\begin{equation}
 \int\int_{\mathcal{E}_{S}}\mathcal{K}dS+\oint_{\partial\mathcal{E}_{S}}kdt
 +\sum_{j}\epsilon_{j}=2\pi\mathcal{Y}(\mathcal{E}_{S}),\label{AH6}
\end{equation}
 and Gaussian curvature is indicated by $\mathcal{K}$ and geodesic curvature is indicated by $k$, declared as
 $k=\bar{g}(\nabla_{\dot{\alpha}}\dot{\alpha},\ddot{\alpha})$ in this way as $\bar{g}
 (\dot{\alpha},\dot{\alpha})=1$, $\ddot{\alpha}$  represents unit acceleration vector and the $\epsilon_{j}$ represents the exterior angle
 at the jth vertex. As $S\rightarrow\infty$, both the jump angles reduce to
 $\pi/2$ and we get $\theta_{O}+\theta_{S}\rightarrow\pi$. The Euler
 characteristic is $\mathcal{Y}(\mathcal{E}_{S})=1$, as $\mathcal{E}_{S}$ is
 non singular. So,
\begin{equation}
 \int\int_{\mathcal{E}_{S}}\mathcal{K}dS+\oint_{\partial
 \mathcal{E}_{S}}kdt+\epsilon_{j}=2\pi\mathcal{Y}(\mathcal{E}_{S}),
\end{equation}
 here, $\epsilon_{j}=\pi$ shows that both
 $\alpha_{\bar{g}}$ and the total jump angle is a geodesic, since the Euler
 characteristic number expressed by $\mathcal{Y}$ is $1$. As
 $S\rightarrow\infty$, the only interesting part to be determined
 is $k(D_{S})=\mid\nabla_{\dot{D}_{S}}\dot{D}_{S}\mid$. Since, radial component of geodesic curvature is defined as \cite{R 35}
\begin{equation}
 (\nabla_{\dot{D}_{S}}\dot{D}_{S})^{r}=\dot{D}^{\phi}_{S}
 \partial_{\phi}\dot{D}^{r}_{S}+\Gamma^{0}_{11}(\dot{D}^{\phi}_{S})^{2}.\label{AH5}
\end{equation}
 For large $S$, $D_{S}:=r(\phi)=S=const$. Hence, the form
 of the Eq. (\ref{AH5}) becomes $(\dot{D}^{\phi}_{S})^{2}
 =\frac{A^2(r)B(r)}{r^2}$. As $\Gamma^{0}_{11}=-rB(r)+\frac{r^{2}A'(r)B(r)}{2A(r)}$, so it becomes
\begin{equation}
 (\nabla_{\dot{D}^{r}_{S}}\dot{D}^{r}_{S})^{r}\rightarrow\frac{1}{S}.
\end{equation}
  As topological defect is not present in geodesic curvature so, $k(D_{S})\rightarrow S^{-1}$. But with the help of optical metric
 Eq. (\ref{AH1}), we can write as $dt=Sd\phi$. Hence, we get this;
\begin{equation}
 k(D_{S})dt=d\phi.
\end{equation}
 With the help of all above results, we obtained
\begin{equation}
 \int\int_{\mathcal{E}_{S}}\mathcal{K}ds+\oint_{\partial \mathcal{E}_{S}} kdt
 =^{S\rightarrow\infty}\int\int_{T_{\infty}}\mathcal{K}dS+\int^{\pi+\Theta}_{0}d\phi.\label{hamza2}
\end{equation}
 Light ray at 0th order in week field limit is defined as $r=b/\sin\phi$, where 'b' is known as impact parameter which is perpendicular distance between path of light and wormhole). Now, using (\ref{AH6}) and (\ref{AH7}), the DAL defined as \cite{R 35}
\begin{equation}
 \Theta=-\int^{\pi}_{0}\int^{\infty}_{b/\sin\phi}\mathcal{K}\sqrt{det\bar{g}} ~dr d\phi.\label{AH7}
\end{equation}

 After putting the leading order terms of Gaussian curvature
 Eq. (\ref{AH1}) into Eq. (\ref{AH7}), the deflection angle is calculated as follows:
\begin{eqnarray}
 \Theta &\thickapprox&  \,{\frac {4a}{3b}}-\,{\frac {\pi\,{a}^{2}}{6{b}^{2}}}+\mathcal{O}  (\frac{a^3}{r^3}).\label{P1}
\end{eqnarray}
Note that for $a=3M$, first order of term of weak deflection angle of Casimir wormhole reduces to the deflection angle of Schwarzschild black hole ($\frac{4 M}{b}$).
\section{Graphical Analysis for non-plasma medium}
 Here, in this section we discuss the graphical behavior of DAL. We observe the relation of deflection angle $\Theta$ with impact parameter $b$ with the variation of $a$. Both 'a' and 'b' have the mass dimension of length so  'L' is the dimension of 'a' and 'b'. In the left graph we change the value of $a$ from $1L$ to $5L$ and in right graph, we vary the values of a from $22L$ to $30L$ by fixing the value of $b$ from $1L$ to $20L$.

\subsection{Deflection angle $\Theta$ w.r.t Impact parameter $b$}
\begin{center}
\epsfig{file=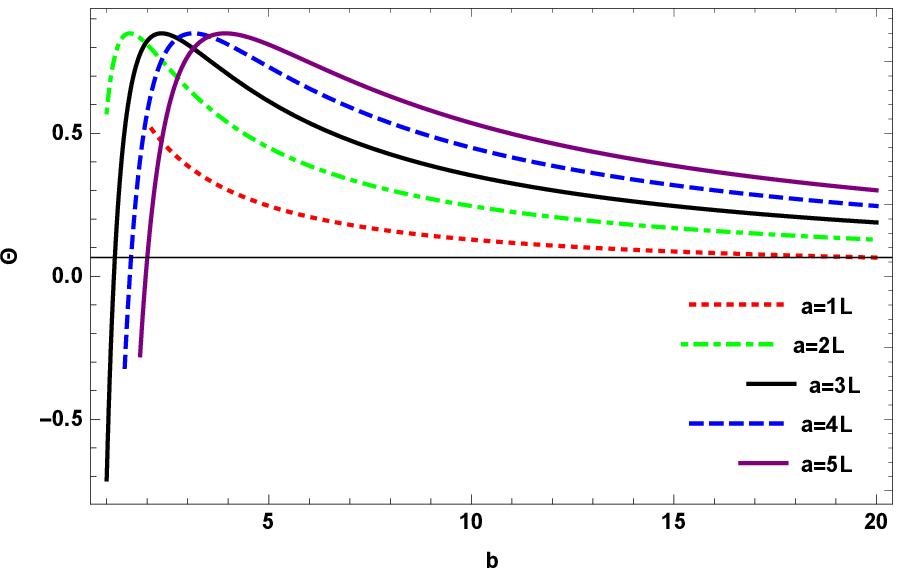,width=0.50\linewidth}\epsfig{file=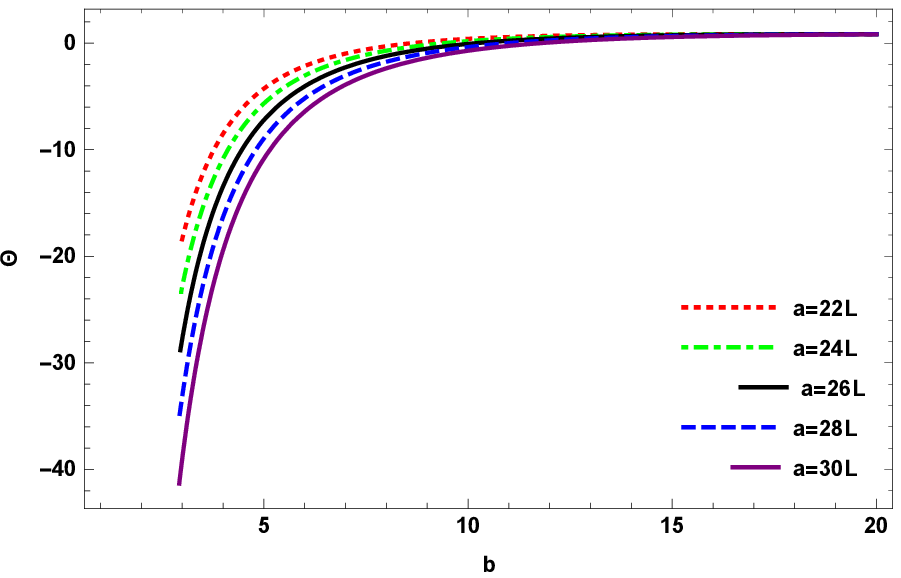,width=0.50\linewidth}\\
{Figure 1: Relation between $\Theta$ and $b$}.
\end{center}
\begin{itemize}
\item \textbf{Figure 1} display the relation of $\Theta$ w.r.t $b$ for different values of $a$.
\begin{enumerate}
\item In left plot, we see that by increasing the values of $a$ the deflection angle first increase toward positive and then gradually decrease from positive to negative that indicates the balanced behavior.
\item In right figure, we observe that for choosing large values of $a$ deflection angle is continuously decreasing from positive to negative.

\end{enumerate}
\end{itemize}
\section{Effect of plasma on gravitational lensing}
 This section is based on the investigation of
DAL by Casimir wormhole in the presence of plasma medium. For
Casimir wormhole the refractive index $n(r)$ \cite{M1 19}, is obtain as,
\begin{equation}
 n^2\left[r,\omega(r)\right]=1-\frac{\omega_e^2(r)}{\omega_\infty^2(r)}.
\end{equation}
 Refractive index is defined as;
\begin{equation}
 n(r)=\sqrt{{1-\frac{\omega_e^2}{\omega_\infty^2}\left(A(r)\right)}}.
\end{equation}
Where $\omega_{e}$ is electron plasma
frequency and $\omega_{\infty}$ is photon frequency evaluated by an observer at infinity and so the corresponding optical metric defined as \cite{M1 19}
\begin{equation}
dt^2=g^{opt}_{lm}dx^ldx^m=n^2 \left[\frac{dr^2}{A(r)B(r)}+\frac{r^2d\phi^2}{A(r)} \right].
\end{equation}
The Gaussian curvature in terms of curvature tensor can be determined as
\begin{equation}
    \mathcal{K}=\frac{R_{r\phi r\phi}(g^{opt})}{det(g^{opt})},\label{hamza1}
\end{equation}
 with the help of Eq. (\ref{hamza1}) Gaussian curvature is written as
\begin{eqnarray}
    \mathcal{K}&\approx& -{\frac {{\omega_e}^{2}a}{{\omega_\infty}^{2}{r}^{3}}}-2/3
\,{\frac {a}{{r}^{3}}}+7/3\,{\frac {{a}^{2}{\omega_e}^{2}}{{r}
^{4}{\omega_\infty}^{2}}}+2/3\,{\frac {{a}^{2}}{{r}^{4}}}.
\end{eqnarray}
 By using GBT, we will calculate the DAL in order to compare it with non-plasma case. In this way, for finding angle in the weak field limits, as the light follows a straight line so we utilize a case of $ r=\frac{b}{sin\phi}$ at 0th order.
\begin{equation}
    \Theta=-\lim_{R\rightarrow 0}\int_{0} ^{\pi} \int_\frac{b}{\sin\phi} ^{R} \mathcal{K} dS.
\end{equation}

 With the help of Eq. (\ref{hamza2}), the DAL in plasma medium is found as follows:
\begin{eqnarray}
\Theta &\thickapprox&  \,{\frac {4a}{3b}}-\,{\frac {\pi\,{a}^{2}}{6{b}^{2}}}-{\frac {7\,\pi\,{a}^{2}{\omega_e}^{2}}{12\,{b}^{2}{\omega_\infty}^{2}}}+2\,{\frac {{\omega_e}^{2}a}{b{\omega_\infty}^{2}}}
 \label{P2}
\end{eqnarray}
The above results tells us that photon rays are moving into medium of
homogeneous plasma. We see that Eq. $(\ref{P2})$ reduced into Eq. $(\ref{P1})$ if plasma effect can be removed.

\section{Graphical Analysis for plasma medium}
In this section, we discuss the effects of plasma medium on the deflection angle of Casimir wormhole by plotting them. We take $\frac{\omega_e}{\omega_\infty}$=$10^{-1}$ and observe behavior of DAL by changing different values of $a$. Here, we also pick the value of $a$ from $1L$ to $5L$ in the left graph and in the right graph, we varry the value of $a$ from $20L$ to $60L$ by fixing the value of $b$ from $1L$ to $20L$.

\subsection{Deflection angle w.r.t Impact parameter}
\begin{center}
\epsfig{file=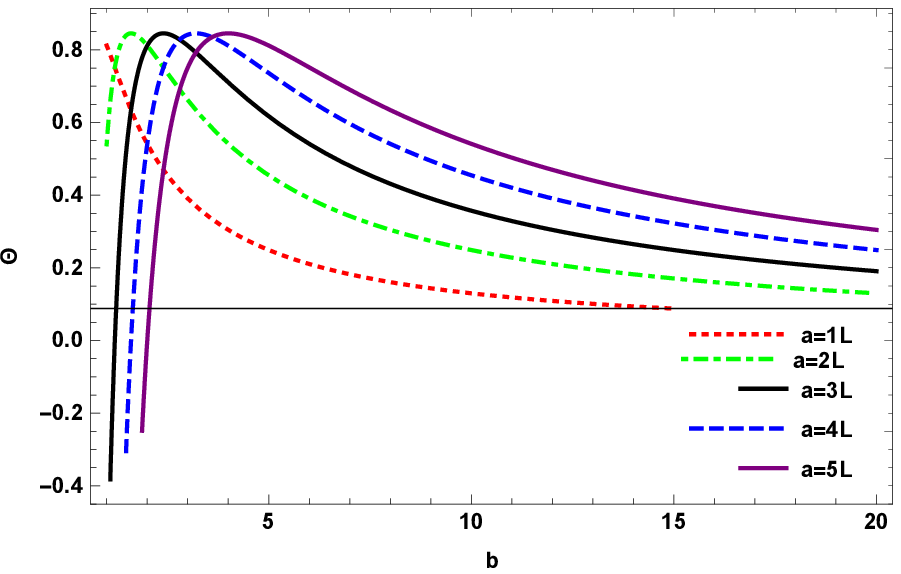,width=0.50\linewidth}\epsfig{file=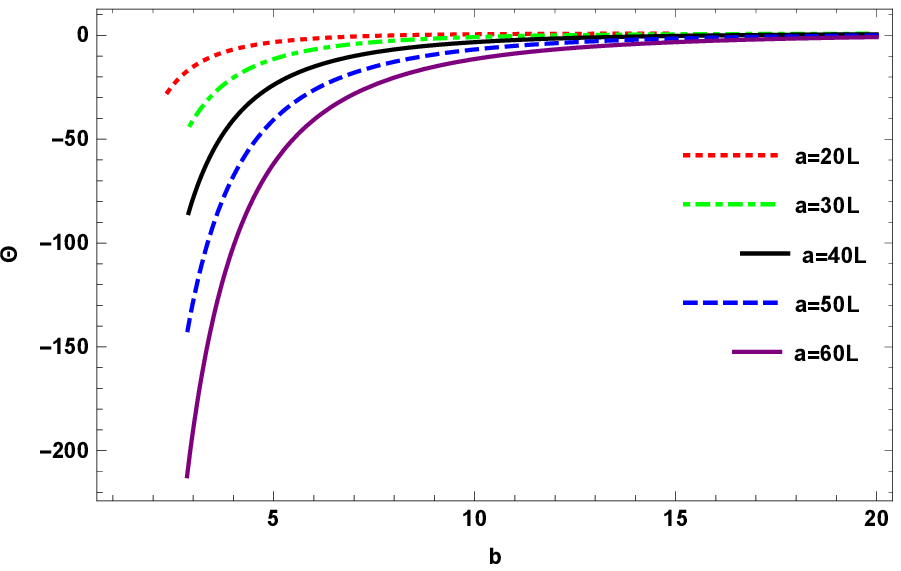,width=0.50\linewidth}\\
{Figure 2: Relation between $\Theta$ and $b$}.
\end{center}

\begin{itemize}
\item \textbf{Figure 2} demonstrates the relation of $\Theta$ with $b$ for different values of $a$.
\begin{enumerate}
\item In 1st plot we show that for small values of $a$ the DAL first increase from negative to positive and then gradually decrease toward negative as like in \textbf{Figure 1}.
\item In 2nd plot we observe that DAL gradually decrease for large values of $a$ as like in
 \textbf{Figure 1}.

\end{enumerate}
\end{itemize}

\section{Null Geodesic in a Casimir wormhole}
The Lagrangian of the Casimir wormhole Eq. (\ref{AH0}) is given by
\begin{equation}
2\mathcal{L}=-\left(1-\frac{2a}{3r}
-\frac{a^2}{3{r^2}}\right) \dot t^2+\left(\frac{1}{1-\frac{2a}{3r}-\frac{a^2}{3{r^2}}}\right)\dot r^2 +r^2\dot \theta^2+r^2\sin^2\theta \dot \phi^2,
\end{equation}
where an overdot means derivative w.r.t the affine parameter $\lambda$. As the Lagrangian is independent of $\phi$ and t so there are only two constants of motion named as the energy $E$ and the angular momentum $L$. Where $E$ and $L$ are defined as:
\begin{equation}
p_{t}=\frac{\partial L}{\partial \dot t}=-(1-\frac{2a}{3r}
-\frac{a^2}{3{r^2}})\dot t=-E
\end{equation}
and
\begin{equation}
p_{\phi}=\frac{\partial L}{\partial \dot \phi}=r^2\sin^2\theta \dot \phi=L.
\end{equation}
Now we can determine the geodesic conditions by utilizing these preserved measures
\begin{equation}
\frac{dt}{d\lambda}=\dot t=\frac{E}{1-\frac{2a}{3r}
-\frac{a^2}{3{r^2}}}~~~,~~~
\frac{d\phi}{d\lambda}=\dot \phi=\frac{L}{r^2\sin^2\theta}.
\end{equation}
The $r$-part and $\theta$-part of the momentum are defined as
\begin{equation}
p_{r}=\frac{\partial L}{\partial \dot r} =\frac{\dot r}{1-\frac{2a}{3r}-\frac{a^2}{3r^2}}~~and~~
p_{\theta}=\frac{\partial L}{\partial \dot \phi}=r^2\dot \theta.
\end{equation}
The $r$-part and $\theta$-part of the geodesic equations can be calculated by working on the Hamilton-Jacobi equation
\begin{equation}
\frac{\partial S}{\partial \lambda}=-\frac{1}{2}g^{\mu \nu}\frac{\partial S}{\partial x^\mu}\frac{\partial S}{\partial x^\nu},\label{S1}
\end{equation}
and for photons $(m_0 = 0)$, Eq. (\ref{S1}) is used by the ansatz
\begin{equation}
S=-Et+L\phi+S_r(r)+S_\theta(\theta),\label{S2}
\end{equation}
here $S_r$ is a function or $r$ and $S_\theta$ is the function of $\theta$.
By putting Eq. (\ref{S2}) into Eq. (\ref{S1}) and also putting the value of contravariant metric, i.e., $g^{\mu\nu}$, and differentiate the terms of variables $r$ and $\theta$ equal to the Carter constant $(\pm \mathcal{K})$ \cite{SA92}, we get
\begin{eqnarray}
\frac{1}{\sqrt{1-\frac{2a}{3r}-\frac{a^2}{3r^2}}}\frac{dr}{d\lambda}=^+_-\sqrt R(r),\nonumber\\
r^2\frac{d\theta}{d\lambda}=^+_-\sqrt T(\theta),
\end{eqnarray}
here $R$ and $\theta$ can be written as
\begin{eqnarray}
R(r)=\frac{E^2}{1-\frac{2a}{3r}
-\frac{a^2}{3{r^2}}}-\frac{\mathcal{K}}{r^2},\nonumber\\
T(\theta)=\mathcal{K}-\frac{L^2}{\sin^2\theta}.
\end{eqnarray}
As we have geodesic equations now our main concern is to find the radial motion of photons across the throat of Casimir wormhole, that can be showed by the computation of the effective potential
\begin{equation}
(\frac{dr}{d\lambda})^2+V_{eff}=0,
\end{equation}
with
\begin{equation}
V_{eff}=-\left({1-\frac{2a}{3r}-\frac{a^2}{3r^2}}\right)R(r).
\end{equation}
From the expression we can say that effective potential depend on constant $a$, radial coordinate $r$ and on energy $E$. It is advantageous to decrease the quantity of parameters by characterizing the impact parameters such as $\xi =\frac{L}{E}$ and $\eta=\frac{\mathcal{K}}{E^2}$. Photon geodesics can be communicated as far as these impact parameters ($\xi$,$\eta$ ). Now we write $\mathcal{R}$, in the form of impact parameters as pursues
\begin{equation}
R=E^2[\frac{1}{1-\frac{2a}{3r}
-\frac{a^2}{3{r^2}}}-\frac{\eta}{r^2}].\label{S3}
\end{equation}
The investigation of effective potential expose the presence of un-stable circular orbits of fixed radius about the throat of wormholes. Those circles are significant from the perspective of optical perceptions.
\section{Shadow of Casimir Wormhole}
Here, we explore the shadow cast of Casimir wormhole. 
First we write conditions to check unstable circular photon orbits:
\begin{equation}
R=0,~~~~~~~~R^{\prime}=0.\label{S4}
\end{equation}
Here prime $(\prime)$ denote derivative w.r.t $r$. Putting (\ref{S3}) into (\ref{S4}), we can easily get
\begin{equation}
\eta=\frac{r^2}{1-\frac{2a}{3r}
-\frac{a^2}{3{r^2}}},~~~~~~~~~~r=\frac{3a^+_-\sqrt{33a^2}}{6}\label{S5}
\end{equation}
 Here, impact parameter $\eta$ depend on radial coordinate $r$ and on constant $a$. Eq. (\ref{S5}) defines the shadow's boundary. Therefore, we make the celestial coordinates ($\alpha$,$\beta$) and associate them with impact parameters ($\xi$,$\eta$). The apparent shape of a shadow are found by using the celestial coordinates (\cite{SA47},\cite{SA48}) as
\begin{eqnarray}
\alpha=^{\lim}_{r_{0} \rightarrow \infty}(r_0^2\sin\theta_0)\frac{d\phi}{dr},\nonumber\\
\beta=^{\lim}_{r_{0} \rightarrow \infty} ~ r_{0}^{2}\frac{d\theta}{dr},\label{S6}
\end{eqnarray}
here distance between wormhole and observer is denoted by $r_0$ and $\theta_0$ is observer's angular coordinate or we can say it "inclination angle". Now putting the equations of four-velocities into Eq. (\ref{S6}), and doing some simple direct calculations we get this type of celestial coordinates
\begin{equation}
\alpha=-\frac{\xi}{\sin\theta_0}~~~~~~~and~~~~~~~ \beta=\sqrt{\eta-\frac{\xi^2}{\sin\theta_{0}^{2}}}.
\end{equation}
Knowing the equations of celestial coordinates and impact parameters, we now make the shadow of Casimir wormhole. So as to construct the shape of shadow, we plot $\alpha$ versus $\beta$ which gives the shadow's boundary. The plots of the shadow for the Casimir wormholes can be saw from Fig 3.

\begin{center}
\epsfig{file=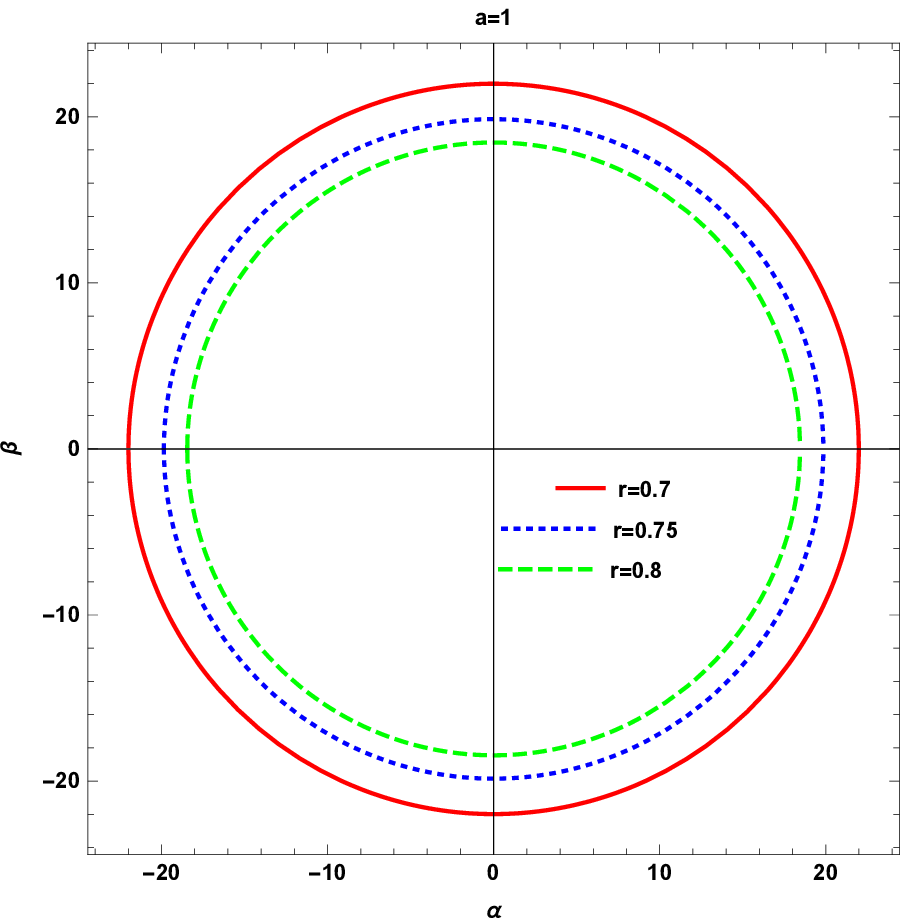,width=0.50\linewidth}\epsfig{file=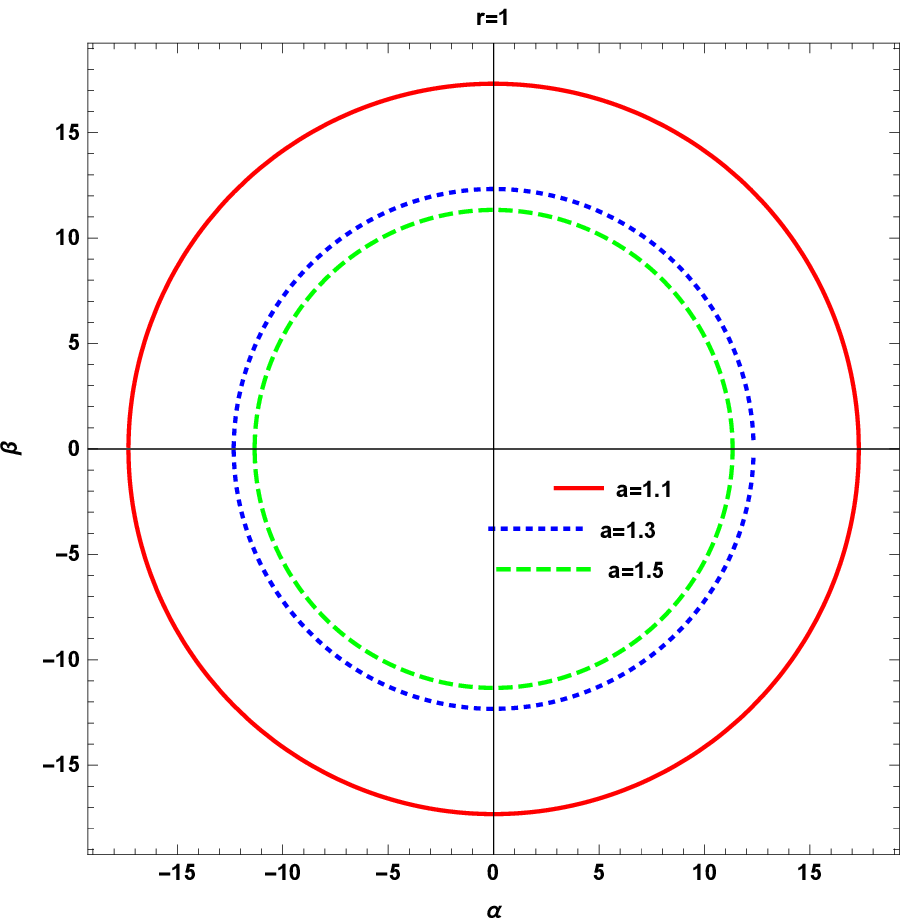,width=0.50\linewidth}\\
{Figure 3: Graphs displaying the shadow of Casimir wormhole}.
\end{center}
By changing the values of constant $(a)$ and radial coordinate $(r)$,
we check the behavior of shadow of Casimir wormhole in the equatorial
plan by putting $\theta_{0}=\frac{\pi}{2}$. We can see from the plots that shadow's shape is a perfect square and it is changed by changing values of
$(a)$ and $(r)$ as shown in Fig. 3. It is seen from the graphs that shadow of
Casimir wormhole reduces by increasing the values of $(a)$ and $(r)$.

\section{Conclusion}
 ~~~~~~In this paper, first of all we have used the GBT to find the DAL by Casimir wormhole. For this purpose, we have obtained the optical Gaussian curvature and have applied the GBT on optical Gaussian curvature to calculate the DAL. The deflection angle has found as follows:
\begin{eqnarray}
 \Theta &\thickapprox&  \,{\frac {4a}{3b}}-\,{\frac {\pi\,{a}^{2}}{6{b}^{2}}}.
\end{eqnarray}
This shows that deflection angle only depends on the impact parameter and the value of $a$. Next we also check the graphical behavior of this deflection angle. We have analyzed the plots that for small values of $a$ deflection angle first decreasing and then goes to positive increase and for large values of $(a)$ deflection angle continuing to decrease from positive to negative side. We have also discussed the deflection angle of Casimir wormhole in the presence of plasma medium. Deflection angle of Casimir wormhole in the presence of plasma medium has obtained as follows:
\begin{eqnarray}
\Theta &\thickapprox&  \,{\frac {4a}{3b}}-\,{\frac {\pi\,{a}^{2}}{6{b}^{2}}}-{\frac {7\,\pi\,{a}^{2}{\omega_e}^{2}}{12\,{b}^{2}{\omega_\infty}^{2}}}+2\,{\frac {{\omega_e}^{2}a}{b{\omega_\infty}^{2}}}.
\end{eqnarray}
Now if we neglect the plasma effect $(\frac{\omega_e}{\omega_\infty}\rightarrow0)$, then this deflection angle Eq. $(\ref{P2})$ reduces into this angle Eq. $(\ref{P1})$. We have also observed the graphical behavior in the presence of plasma medium. In these plots we have observed that the deflection angle in plasma medium is increased as compared with non-plasma medium. \\

Then, we have calculated the shadow of Casimir wormhole. To do so, we have first obtained the null geodesic of the Casimir wormhole and then we have found the apparent shape of a shadow by using the celestial coordinates. It has been seen in plots that its shadow is a perfect circle and this circle change by changing values of $r$ and $a$. Hence, the graphs of the shadows show that the radius of the shadow decreases when we increase the value of $r$ and $a$.

\end{document}